4

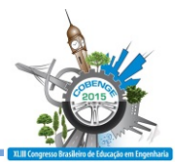

# APRENDENDO PROGRAMAÇÃO ORIENTADA A OBJETOS COM UMA ABORDAGEM LÚDICA BASEADA EM GREENFOOT E ROBOCODE

**Clêison Simões dos Santos** – cleison.ecomp@gmail.com
**Allen Hichard Marques dos Santos** – allenhichard21@gmail.com
**Suenny Mascarenhas Souza** – suenny.s@gmail.com
**David Moises B. dos Santos** – davidmbs@uefs.br
**Roberto Almeida Bittencourt** – roberto@uefs.br
Universidade Estadual de Feira de Santana
Avenida Transnordestina, s/n – Novo Horizonte
44.036-900– Feira de Santana – Bahia

*Resumo: Facilitar a aprendizagem de programação orientada a objetos (POO) é um dos grandes desafios em cursos de computação, pois este paradigma possui uma variedade de conceitos com nível de abstração considerado elevado para iniciantes, mesmo para os que já programam num paradigma imperativo. Além disto, a transição do paradigma imperativo para o orientado a objetos é complexa, e normalmente apresenta efeitos colaterais indesejados. Um esforço significativo vem sendo direcionado na busca de soluções motivadoras e atrativas para estas questões. Uma delas é o uso de ambientes lúdicos, que mesclam o uso de jogos com aprendizagem. Neste trabalho, relatamos nossa experiência com oficinas de aprendizagem de POO através de jogos, desafios e competições apoiadas pelas ferramentas Greenfoot e Robocode. Uma oficina com estudantes do segundo período de um curso de Engenharia de Computação é apresentada aqui. A experiência evidenciou lições importantes para motivar os estudantes: elaboração de bons exemplos, o uso de desafios competitivos e a boa razão entre tutores/monitores e participantes. Além disso, a oficina mostrou-se prática e efetiva para introduzir e motivar os estudantes a aprender POO.*

*Palavras-chave: Aprendizagem ativa, programação orientada a objetos, jogos, Greenfoot, Robocode.*

## 1. INTRODUÇÃO

A aprendizagem de programação orientada a objetos (POO) tem sido discutida pela comunidade acadêmica há cerca de duas décadas (KÖLLING; ROSENBERG, 1996). Estes autores sugerem que, apesar da elegância do paradigma orientado a objetos e suas vantagens associadas, sua introdução em escala na indústria de tecnologia de informação levou à necessidade de ensinar conceitos bastante abstratos a aprendizes iniciantes de programação. Questões como encapsulamento, herança, interfaces versus implementação, polimorfismo, antes ensinadas em tópicos avançados, são agora elementos básicos de disciplinas introdutórias à programação. Além disso, é muito comum em muitas universidades a introdução do paradigma de programação imperativo antes do orientado a objetos, o que leva muitos estudantes a escreverem código parecido com imperativo em linguagens orientadas a objetos, gerando inúmeros outros problemas (KÖLLING; ROSENBERG, 1996). Evidências apontam que estudantes levam, em média, entre 6 a 18 meses para transitar entre os dois paradigmas (KÖLLING, 1999). Para os estudantes, estes novos conceitos abstratos e a confusão entre os dois paradigmas faz com que a abordagem tradicional de ensino-aprendizagem não seja suficiente para introduzir POO e manter os estudantes motivados (BEGOSSO *et al.*, 2012). Além de as linguagens de programação serem muito complexas, há o problema adicional de os ambientes de desenvolvimento serem muito confusos para iniciantes (KÖLLING, 1999).

Kölling sugere alguns elementos importantes para uma linguagem de programação orientada a objetos para iniciantes. Ela deve ser simples e fácil de compreender e deve ser puramente orientada a objetos. Os conceitos aprendidos em uma linguagem devem ser facilmente transferidos para qualquer outra linguagem. Além disso, a linguagem deve fornecer um bom ambiente de desenvolvimento para apoiá-la, escondendo detalhes desnecessários, e permitindo aos programadores se concentrarem apenas na programação em si (KÖLLING, 1999).

Estudantes do terceiro e quarto anos de computação da Universidade de Guelph-Humber revelaram preferir um ambiente de programação mais interativo e divertido: uma vez entretidos, eles se tornam mais interessados no que estão fazendo (GALLANT; MAHMOUD, 2008). Os autores sugerem que, para extrair o melhor dos estudantes, eles devem ser desafiados, testados e entretidos.

Como resposta a este cenário, alguns ambientes foram desenvolvidos para facilitar a aprendizagem de POO. Ferramentas como *Greenfoot* (KÖLLING, 2010), *BlueJ* (KÖLLING *et al.*, 2003), *Alice* (COOPER, 2010) e *Robocode* (HARTNESS, 2004), por exemplo, buscam fornecer um ambiente de desenvolvimento lúdico, onde a programação se torna mais simples e divertida para os iniciantes em POO.

Associadas a ferramentas para iniciantes, projetos têm sido realizados, utilizando uma metodologia de ensino baseada na construção de jogos e animações. Buscando melhorar a experiência dos estudantes em programação, laboratórios de programação com a ferramenta *Greenfoot* foram realizados para a construção de um projeto chamado *Going to the Moon* (GALLANT; MAHMOUD, 2008). Outro projeto similar utilizou a ferramenta *BlueJ* para desenvolver o jogo de cartas 21 (*Black Jack*), como proposta para reduzir as dificuldades apresentadas por estudantes ao aprender POO (KOUZNETSOVA, 2007). Para a autora, *BlueJ* e o jogo fornecem um ambiente familiar e divertido que aumenta até mesmo o nível de envolvimento dos alunos.

Inspirados por essas e outras experiências, nosso grupo de pesquisa e extensão, realiza, há cerca de dois anos, experiências alternativas de ensino-aprendizagem de programação, inclusive de POO. Este artigo relata nossa experiência de oficinas de POO com ferramentas lúdicas, descrevendo em detalhes a última oferta destas oficinas no período 2015.1 do curso de Engenharia de Computação da Universidade Estadual de Feira de Santana (UEFS). Estas oficinas objetivam reduzir as dificuldades apresentadas pelos alunos ao iniciarem o aprendizado de POO, bem como avaliar e melhorar a abordagem utilizada. Esta última oficina utilizou as ferramentas *Greenfoot* e *Robocode* e uma abordagem que combina desenvolvimentos de jogos com desafios propostos aos estudantes.

Este artigo está organizado como a seguir. A Seção 1 é dedicada a revisar trabalhos relacionados, seguindo pela Seção 2 com a metodologia utilizada neste trabalho. Os resultados são descritos na Seção 3 e as lições aprendidas na Seção 4. Por fim, expomos, na Seção 5, nossas conclusões e sugestões de trabalhos futuros.

## 2. TRABALHOS RELACIONADOS

A aprendizagem de programação é um tema envolto de complexidades e dificuldades. É comum os estudantes terem dificuldade em compreender as abstrações envolvidas na programação de computadores, principalmente os programadores iniciantes em cursos de graduação em computação. Em POO, as dificuldades são geralmente relacionadas aos conceitos inerentes ao paradigma orientado a objetos, às linguagens utilizadas e aos ambientes de desenvolvimento utilizados para a aprendizagem (KÖLLING, 1999). Pesquisadores buscam saídas que mantenham os estudantes motivados, e assim diminuam os índices de reprovação e evasão nestes cursos (BEGOSSO *et al.*, 2012). Como resultado destas pesquisas, ambientes têm sido desenvolvidos especificamente para iniciantes (KELLEHER; PAUSCH, 2005), e o foco de vários deles têm sido em permitir que seus usuários trabalhem em temas ligados a seus próprios interesses, como é o caso de jogos e animações (GUZDIAL, 2004). *Greenfoot* é um destes ambientes, voltado especificamente para facilitar a aprendizagem de POO (KÖLLING, 2010). Um outro ambiente voltado para POO e baseado em competições de robôs é o *Robocode* (HARTNESS, 2004).

*Greeenfoot* tem sido utilizado de várias maneiras por pesquisadores. A abordagem feita por Gallant e Mahmoud (2008) consiste em aplicar a ferramenta em laboratório, dividindo em 10 etapas durante o semestre, com duração de duas horas cada. Assim, em cada sessão, há um dado conjunto de conceitos a serem trabalhados, acompanhados de exemplos no *Greenfoot*. Este trabalho reporta ainda que os estudantes prefeririam se divertir fazendo programas para um ambiente gráfico do que fazer programas com interfaces de linha de comando.

Begosso e colaboradores (2012) ministraram um curso divido em uma parte teórica e outra prática. Iniciaram o curso com os conceitos principais desse paradigma. O fluxo continuou com aulas práticas de *Greenfoot* utilizando exemplos de jogos em uma sequência que facilitasse a aprendizagem. O primeiro exemplo visava apresentar a interface do *Greenfoot* para invocar métodos, tipos de retorno e parâmetros. Conceitos de herança e edição de código foram praticados na segunda etapa. No terceiro exemplo alunos se debruçaram sobre a edição de código-fonte, assinatura de métodos, recursão e compilação de classes. Por fim, uma avaliação sobre conhecimentos de POO adquiridos foi aplicada aos participantes. O resultado mostrou que maioria dos estudantes aprendeu os conceitos apresentados.

A proposta do trabalho de Montero et al. (2010), é, em resumo, testar o uso de *Greenfoot* e *jGRASP* (uma ferramenta para visualização de programas em execução) para auxiliar no ensino-aprendizagem de POO. Um grupo controle com 15 alunos e outro grupo experimental com 18 alunos foram criados, ambos com conhecimento similar comprovado por nivelamentos e testes. O experimento teve duas sessões de implementação das atividades, onde o grupo experimental usou *jGRASP* e *Greenfoot* e o grupo controle, apenas *jGRASP*. Com um questionário pós-experimento, o grupo experimental apresentou pontuação média acima do grupo controle (MONTERO *et al.*, 2010).

O trabalho de Liu (2008) descreve uma experiência de utilização de *Robocode* durante três semanas com 26 estudantes de graduação. Relata que o uso de *Robocode* retrata situações realísticas, o que motiva os participantes. Também reporta que a necessidade exploratória e de autoaprendizagem é bastante presente em relatos feitos pelos estudantes.

Uma outra abordagem fez uso da aprendizagem baseada em problemas (do inglês, PBL), aplicada para estudantes de graduação, onde o objetivo era desenvolver batalhas com os robôs implementados pelos estudantes de forma a retratar a realidade em um âmbito profissional competitivo (O'KELLY; GIBSON, 2006). Para tanto, foram montadas equipes, visto que é comum a programação em pares. Os vencedores participam de uma batalha em esfera nacional, motivando o espírito competitivo dos estudantes.

## 3. METODOLOGIA

Esta seção descreve as ferramentas usadas, os participantes, o planejamento da oficina e os procedimentos para coleta e análise de dados.

### 3.1. As ferramentas

***Greenfoot*** é um ambiente de desenvolvimento integrado de software voltado para o ensino-aprendizagem de POO para jovens a partir de 14 anos. Reúne ferramentas como editor de código, compilador, máquina virtual, além de ferramentas educacionais (KÖLLING, 2008). Usa Java como linguagem de programação, embora permita o uso da linguagem de forma mais simples (KÖLLING, 2010). Conceitos importantes de POO tais como design de classes, separação de responsabilidades, encapsulamento, coesão, acoplamento e níveis adicionais de abstração podem ser explorados neste ambiente.

***Robocode*** é um ambiente de desenvolvimento integrado para aprendizagem de programação através de programação orientada a eventos. Desenvolvido em 2001 por Mathew A. Nelson na IBM, tornou-se *open source* em 2005. O ambiente é baseado em um micromundo virtual onde cada aprendiz desenvolve uma classe de um tanque de guerra em Java para combater outros tanques em uma arena. As batalhas são executadas em tempo real e com representação gráfica.

### 3.2. Participantes

A oficina teve a participação de 15 alunos do curso de Engenharia de Computação da UEFS: 13 do sexo masculino e duas do sexo feminino. Todos os participantes tinham experiência de pelo menos um semestre com programação imperativa na linguagem C, e iriam cursar o componente curricular de POO logo após a oficina, normalmente oferecido no segundo semestre de curso.

### 3.3. Planejamento da oficina

O objetivo da oficina é que seus participantes sejam capazes de escrever pequenos programas orientados a objetos nos ambientes Greenfoot e Robocode, compreendendo conceitos básicos de POO. Um objetivo colateral é o de reduzir os índices de evasão e reprovação das disciplinas de POO.

A oficina de POO ocorreu no campus da UEFS e foi guiada por dois tutores e três monitores. A Tabela 1 dá uma visão geral da organização da oficina.

Tabela 1 – Visão geral da oficina de POO com Greenfoot e Robocode.

| Oficinas de Programação Orientada a Objetos com Greenfoot e Robocode | |
|---|---|
| **Objetivos** | Ser capaz de desenvolver pequenos programas orientados a objetos nos ambientes Greenfoot e Robocode. |
| **Metodologia** | Aprendizagem através de descoberta no ambiente Scratch, guiada por desafios propostos pelos tutores, em especial, o desenvolvimento de jogos. |
| **Conteúdo** | Classes e objetos. Atributos, métodos e mensagens. Herança, generalização e especialização. Sobrescrita de métodos e polimorfismo. |
| **Tutoria** | Dois tutores e três monitores, todos estudantes do curso de Engenharia de Engenharia de Computação que cursaram previamente as disciplinas de programação orientada a objetos. |
| **Participantes** | 15 estudantes do segundo semestre do curso de Engenharia de Computação da UEFS. |
| **Local** | Laboratório de Programação de Engenharia de Computação, Campus da UEFS. |
| **Período** | 23 a 27 de fevereiro de 2015, turno vespertino. |
| **Carga Horária** | 20 horas, divididas em cinco sessões de quatro horas. |

A abordagem de ensino-aprendizagem foi uma combinação de pequenos tutoriais sobre conceitos de POO, exploração das ferramentas por descoberta guiada, além de usar desafios na forma de jogos e competições. Os tutores e monitores propõem desafios graduais para os participantes, e eles os solucionam utilizando um dos ambientes lúdicos. Dúvidas ao longo do processo são sanadas pelos tutores ou monitores. A aprendizagem é centrada no estudante, incentivando que cada um busque, no seu próprio tempo, seu modo particular de resolver os problemas de desenvolvimento – os monitores só intervêm quando é realmente necessário, para tirar dúvidas ou evitar bloqueios.

A oficina foi realizada em cinco dias. A Tabela 2 detalha o planejamento de cada dia da oficina, com as atividades desenvolvidas e o conteúdo associado. Os três primeiros dias são dedicados a *Greenfoot* e os últimos dois dias, a *Robocode*. Nestes dois últimos dias nenhum conteúdo novo de orientação a objetos é apresentado, pois a finalidade é, através de outro projeto, reforçar alguns conceitos trabalhados anteriormente. No final terceiro dia de *Greenfoot,* apresenta-se o *NetBeans*, como uma forma de transição entre um ambiente educacional e outro profissional. Neste caso, usa-se um exemplo com um simples sistema de informação.

Tabela 2 – Planejamento da oficina de POO com Greenfoot e Robocode.

| # | Projeto | Atividade | Conteúdo |
|---|---------|-----------|----------|
| 1 | *Wombat* | Apresentação do ambiente *Greenfoot* em paralelo a conceitos básicos de POO. Manipulação do *Greenfoot* através do jogo *Wombat*. | Objetivos do *Greenfoot*, ambiente do *Greenfoot*. Objetos e Classes. Atributos, métodos e mensagens. Herança e Polimorfismo. |
| 2 | *Striker Gunner* | Criação de jogo de guerra espacial. Importação de mídias. Movimentação do cenário. Exploração da API Java do framework. Movimentação da nave do jogador e de nave inimiga. Início da interação entre naves. | Herança. Criação de subclasses de Mundo e Ator. Construtores. Instanciação de objetos. Alternar imagens dos objetos. Leitura do teclado. Criação de métodos. Parâmetros dos métodos e tipo de retorno. Interação entre objetos. |
| 3 | *Striker Gunner* / Sistema de Informação | Aparecimento aleatório de naves inimigas. Pontuação do jogo. Colisão entre naves, entre nave e a tiros. Detecção de fim de jogo. Introdução ao ambiente NetBeans com visualização e modificação de um pequeno sistema de informação. | Chamadas de métodos de classes do framework. Números aleatórios. Variáveis. Listas encadeadas. Leitura de por outras classes através de métodos. Acesso a estado a partir de classes diferentes. Ambiente NetBeans. Leitura e criação de classes no NetBeans, evidenciando semelhanças com Greenfoot. |
| 4 | Robocode | Ambientação com a ferramenta: exploração do código Java dos robôs e criação de uma batalha. Discussão do código dos robôs. Implementação de um robô pelos alunos. | Leitura de classes. Instanciação de objetos. Elaboração de métodos em Java. Sobrescrita de métodos em Java. |
| 5 | Robocode | Conclusão do código dos robôs. Realização das batalhas de robôs. | Elaboração de métodos. Variáveis. Mensagem entre objetos. |

### 3.4. Coleta e análise de dados

Coletamos dados das oficinas tanto quantitativos, na forma de questionários, como qualitativos, na forma de reflexões dos tutores, observações e entrevistas com os alunos. Cada aluno recebeu um termo de consentimento livre e esclarecido informando sobre os objetivos e procedimentos usados neste trabalho, e todos aceitaram participar. Os resultados e lições aprendidas deste trabalho estão apoiados nas reflexões e em algumas questões do questionário, analisadas através de estatística descritiva. Reflexões, observações e entrevistas ainda estão em análise e serão discutidas oportunamente em outro trabalho.

## 4. RESULTADOS

Os resultados são descritos a seguir, usando uma narrativa em ordem cronológica, ressaltando fatos, percepções e comportamentos, complementada com uma análise mais geral.

**Primeiro dia** – Os tutores apresentaram o ambiente Greenfoot em paralelo com noções de orientação a objetos. Ocorreu uma dinâmica participativa, através de explanações, questionamentos e discussões. Após uma navegação sobre as funcionalidades da ferramenta, foram trabalhados os conceitos de POO através do projeto *Wombat* (ver Figura 1). Estudantes exploraram os objetos através dos menus de contexto. Para fundamentar mais os conceitos de classes, objetos e atributos, foram apresentados exemplos do dia-a-dia, seguidos de demonstração com a ferramenta. Por conta dos questionamentos dos estudantes, foram discutidas algumas palavras-chave da linguagem Java. O acesso à documentação do framework através do ambiente foi então explorado pelos participantes através das classes *Ator*, *Mundo* e *Greenfoot* e de seus métodos. Isto os levou a tentar resolver sozinhos os problemas relacionados à implementação do jogo, reduzindo a necessidade de intervenção dos tutores e monitores. Um desafio final do dia é lançado para modificar o *Wombat*, permitindo movimentar os objetos através das setas do teclado.

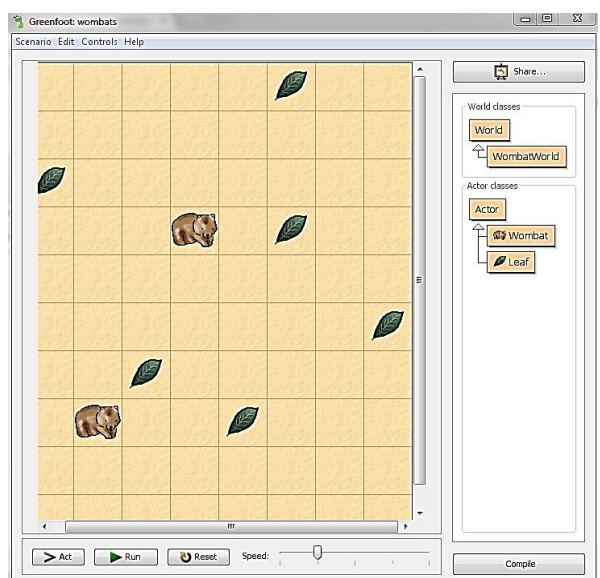

Figura 1 – Cenário do projeto Wombat.

**Segundo dia** – Os tutores propuseram a criação de um jogo de batalha espacial chamado *Striker Gunner*. Algumas orientações básicas sobre o jogo foram dadas, assim como dicas para interação com o Greenfoot de modo a colocar objetos no cenário (nave, inimigos, plano de fundo), movimentar objetos, e lançar tiros. Muitos sentiram dificuldades para iniciar, recorrendo ao exemplo do primeiro dia para revisão. Vários estudantes apresentaram dificuldades em elaborar seus próprios métodos, pela sintaxe da declaração de métodos e/ou pela chamada de métodos das classes *Mundo* e *Ator*. Por outro lado, todos tiveram facilidade em fazer o movimento da *Nave*. Outra dificuldade surgiu na elaboração da lógica para o movimento contínuo do plano de fundo. Todos os monitores e tutores foram requisitados na ocasião, além da colaboração entre os próprios participantes. Os estudantes trabalharam os conceitos de classes, atributos e métodos, herança e polimorfismo.

**Terceiro dia** – Após boa parte dos estudantes serem introduzidos a POO, implementando funcionalidades do jogo, foram propostos desafios para a conclusão do jogo: a inserção de mais naves inimigas no cenário, implementação de colisões entre naves e a definição de estados de vitória ou derrota. Vários estudantes apresentaram dificuldades em implementar os tiros, especialmente ao tentar usar instâncias que não estavam mais presentes no *Mundo*. Com suporte dos monitores, esses impasses foram resolvidos. Ao verem o resultado final, muitos alunos demonstram contentamento através de expressões de animação.

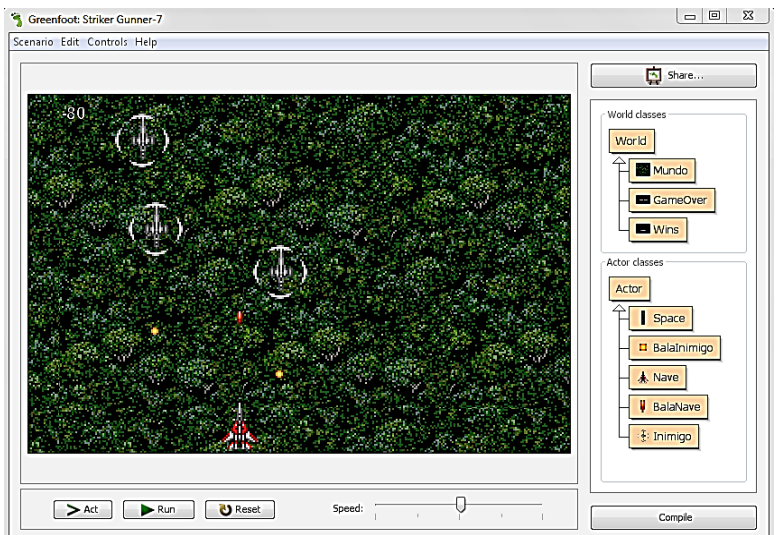

Figura 2 – Jogo *Striker Gunner* implementado por um participante.

Os participantes que concluíram mais cedo passaram a adicionar mais efeitos nos seus jogos, não exigidos nos desafios. Após o término desta etapa, passa-se a uma introdução ao NetBeans. Os estudantes lêem e modificam uma aplicação Java de cadastro de alunos, professores e disciplinas, tentando colocar em prática os conhecimentos adquiridos. As dúvidas mais comuns foram em relação ao uso de construtores, já que no *Greenfoot* eles não tinham esta preocupação, e no uso de listas encadeadas para organização dos cadastros.

**Quarto dia** – Inicia-se uma nova etapa da oficina cujo desafio foi uma competição de robôs feita em Java através da ferramenta *Robocode*. Os tutores apresentaram a ferramenta, alguns códigos existentes de criação do robô, a API, e as regras para a disputa. Em seguida, os estudantes formaram duplas. Cada uma delas deveria fazer o seu próprio Robô para participar da batalha. Cada dupla elaborou livremente sua estratégia a partir dos métodos encontrados na documentação e do questionamento constante aos tutores e monitores.

**Quinto dia** – As equipes tiveram um tempo adicional para terminar a implementação dos robôs. Em seguida, houve a disputa entre os robôs das equipes. Todos os robôs foram colocados na arena e, após 10 rodadas de jogo, foi escolhido o melhor robô. A equipe que desenvolveu o melhor robô recebeu um prêmio como lembrança da oficina.

Em uma análise mais geral, percebemos participação integral na oficina, com apenas um aluno não cumprindo a carga horária total. Os participantes avaliaram a oficina através de um questionário. De modo geral, a avaliação foi positiva. Seis diferentes aspectos foram examinados em uma escala de Likert, variando de zero para o mais insatisfatório até cinco para o mais satisfatório. A Figura 3 apresenta as médias das avaliações de cada dimensão. Em geral, os resultados foram bastante positivos, com média acima de 4, exceto o aspecto cansativa/leve, talvez por a oficina ser bastante intensiva.

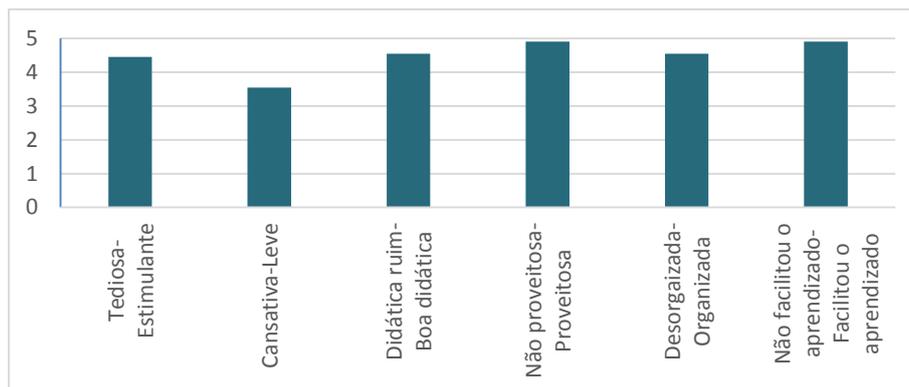

Figura 3 – Avaliação da oficina

## 5. LIÇÕES APRENDIDAS

É importante sintetizar as principais lições desta experiência, que consideramos: exemplos motivantes, atenção dividida, boa relação entre tutores/monitores e alunos, transição para um IDE profissional e o uso de desafios e competições.

*Exemplos motivantes.* Uma oficina intensiva, com vinte horas de duração em uma semana, precisa manter a atenção constante. Isto realmente aconteceu na maior parte do tempo. Alguns exemplos utilizados foram divertidos, gerando momentos de descontração na oficina, e mantendo um clima leve e divertido.

*Atenção dividida.* Em alguns momentos, mesmo usando uma abordagem ativa de aprendizagem, alguns alunos se dispersaram, especialmente ao ser-lhes dados algum tempo para analisar e modificar sozinhos o código-fonte. Quando todos os monitores e tutores estavam ocupados, alguns alunos "bloquearam" e acabaram se dispersando da oficina. Alguns passaram a jogar no celular ou mesmo acessar a Internet para outros fins. Obviamente, isso depende do interesse de cada participante, mas é um elemento importante a ser levado em conta. Ter um maior número de monitores disponíveis fez com que os alunos se sentissem mais seguros, podendo contar com ajuda constante para tirar dúvidas sempre que necessário, o que diluiu bastante estas situações de dispersão.

*Boa relação entre tutores/monitores e alunos.* Alguns exemplos utilizados foram divertidos, gerando momentos de descontração na oficina e mantendo um clima leve e divertido. O fato de os tutores e monitores serem pares dos participantes contribuiu, sem dúvida, para isso.

*Transição para um IDE profissional.* Nesta oficina, experimentamos uma transição para o IDE NetBeans, que vários estudantes utilizam nas disciplinas de POO na UEFS. Há algum tempo, percebemos que, mesmo após o uso do *Greenfoot*, algumas dúvidas são frequentes quando os participantes implementam uma aplicação Java num IDE profissional. Portanto, ter acesso a uma transição com o ambiente *NetBeans* foi bastante positiva, pois eles conseguiram perceber como os conceitos de POO se materializam num ambiente profissional.

*Desafios e competições.* Outro elemento positivo das oficinas foi a competição feita no *Robocode*. Embora alguns tutores e monitores tivessem dúvidas sobre a eficácia, pela simplicidade do micromundo do ambiente, houve mais adesão e menos dispersão em comparação às aulas com *Greenfoot*. Os alunos mergulharam na produção de seus robôs e, na disputa, todos ficaram bastante envolvidos. Isto foi, com certeza, um dos fatores que levou a maioria dos alunos a gostar mais da ferramenta *Robocode* do que do *Greenfoot*.

## 6. CONCLUSÕES

Este artigo relatou uma experiência de oficinas lúdicas de POO através de uma abordagem de ensino-aprendizagem ativa com uso de desafios, jogos e competições e com o apoio dos ambientes de programação para iniciantes *Greenfoot* e *Robocode*. As oficinas foram oferecidas a estudantes de Engenharia de Computação do segundo semestre da UEFS, em um período de uma semana antes do início do semestre letivo, com carga horária de vinte horas.

Os resultados sugerem que, usados de forma adequada, jogos, competições e desafios permitem aumentar a motivação para o aprendizado, e tornar o processo de aprendizagem leve e agradável. Percebemos ainda que o uso de competições deixam as atividades mais excitantes para os estudantes, levando-os a quebrar suas próprias barreiras no processo de aprendizagem. Um bom planejamento, com desafios e exemplos motivantes, e com uma relação adequada de tutores/monitores por aluno deve ser também levado em conta para o sucesso da experiência. Concluímos que a oficina realizada funcionou realmente como instrumento motivador e introdutório para o aprendizado de POO, e que este tipo de esforço tem o seu espaço nos cursos de graduação em computação.

Em trabalhos futuros, pretendemos analisar pormenorizadamente os dados obtidos de questionários, entrevistas e observações feitas em um período mais longo de oficinas de POO. Pretendemos ainda continuar com esta abordagem de oficinas antes das aulas, por terem mostrado papel importante na eliminação das barreiras iniciais à aprendizagem de POO. Finalmente, pretendemos ainda experimentar com outras combinações de ferramentas e atividades, evoluindo as oficinas para um curso introdutório lúdico de POO a ser oferecido em escolas de ensino médio.

### *Agradecimentos*

## REFERÊNCIAS BIBLIOGRÁFICAS

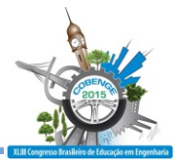

# LEARNING OBJECT-ORIENTED PROGRAMMING WITH A PLAYFUL APPROACH BASED ON GREENFOOT AND ROBOCODE

*Abstract: One the major challenges in undergraduate computing programs is the learning of object-oriented programming (OOP). This paradigm has a variety of concepts with an abstraction level usually high for most beginners, even the ones who already code in an imperative language. Furthermore, transitioning from imperative programming to OOP is a complex issue, with various inappropriate side effects. A significant effort has been pursued in the search of motivating and attractive solutions for such issues. One of those is the use of playful environments that merge games with learning. In this work, we report our experience with OOP learning workshops by means of games, challenges and competitions, supported by Greenfoot and Robocode learning environments. A workshop with sophomore students in a Computer Engineering program is presented here. Lessons learning to motive students include: design of motivating examples, use of competitive challenges, and an appropriate ratio between tutors and students. Results suggest that the workshop was a practical and effective way to introduce OOP and motivate students to learn it.*

***Key-words:*** *active learning, object-oriented programming, games, Greenfoot, Robocode.*